\documentclass[final,3p,times,preprint]{elsarticle}

\usepackage{times,amsmath,epsfig}
\usepackage{subfigure}
\usepackage{cite}
\usepackage{comment}
\usepackage{amssymb}
\usepackage{graphicx}
\usepackage{url}
\usepackage{balance}
\usepackage{amsmath} 
\usepackage{color}
\usepackage{multirow}
\usepackage{afterpage}
\usepackage{algorithm}
\usepackage{verbatim}
\usepackage[noend]{algpseudocode}
\usepackage[table,xcdraw]{xcolor}

\newfloat{program}{thp}{lop} 
\floatname{program}{Program}
\newcommand{\ie}{i.e.} 
\newcommand{\eg}{e.g.} 
\newcommand{\et}{et al. }

\journal{Energy Journal}

\begin{document}

\begin{frontmatter}



\title{Regression-based Online Anomaly Detection for Smart Grid Data}




\author{Xiufeng Liu}
\cortext[cor1]{Corresponding author}
\ead{xiuli@dtu.dk}

\author{Per Sieverts Nielsen}

\address{Technical University of Denmark}

\begin{abstract}
With the widely used smart meters in the energy sector, anomaly detection becomes a crucial mean to study the unusual consumption behaviors of customers, and to discover unexpected events of using energy promptly. Detecting consumption anomalies is, essentially, a real-time big data analytics problem, which does data mining on a large amount of parallel data streams from smart meters. In this paper, we propose a supervised learning and statistical-based anomaly detection method, and implement a {\em Lambda} system using the in-memory distributed computing framework, {\em Spark} and its extension {\em Spark Streaming}. The system supports not only iterative detection model refreshment from scalable data sets, but also real-time detection on scalable live data streams. This paper empirically evaluates the system and the detection algorithm, and the results show the effectiveness and the scalability of the proposed lambda detection system.
\end{abstract}


\end{frontmatter}

\section{Introduction}
\label{sec:intr}
 Anomaly detection, also known as outlier detection, is the process of discovering patterns in a given data set that do not conform to expected behavior \citep{Akyildiz}. Anomaly detection is to find the events that happen relatively infrequently, which has been extensively used in a wide variety of applications, including fraud detection for credit cards, insurance, health care, intrusion detection for cyber-security, fault detection in safety critical systems, and many others \citep{Akyildiz}. Smart meter analytics attracts the growing research effort \eg, \citep{sdewes2015,icde2015,energy2016,edbt2015}, due to the wide installation of smart meters. Anomaly detection can be applied for analyze live smart meter data, which aims to help energy consumers identify unusual behaviors, e.g., forgetting to turn off stoves after cooking; and to help utilities detect extraordinary events, e.g., energy leakage and theft. Since abnormal consumption may also result from user activities, such as using inefficient appliances, or over-lighting and working overtime in office buildings, anomalous feedback can be used to warn energy consumers to minimize energy usage and to help them identify inefficient appliances or over-lighting. Furthermore, anomaly detection can help utilities to establish the baseline for providing more accurate demand-response programs to their customers \citep{zhang2011}. Abnormal energy consumption detection is related to finding patterns in data, and the statistical and data mining techniques are used to detect the patterns, e.g., \citep{zhang2011,nadai2015,liu2011,Chou2014}, which can perform close to or better than domain experts. 
 
Today, smart meters are widely used worldwide. Smart meters are the digital devices that can record energy consumption at the interval of an hour or fewer \citep{Depuru2011}. Smart meters record detailed consumption readings in real-time or near real-time, which provide the opportunity to monitor timely unusual events or consumption behaviors. However, the enabling anomaly detection for smart meters typically uses data mining technologies, which require large amounts of training data sets, as well as significantly complex systems. In typical applications of data mining to anomaly detection, the detection models are produced off-line because the learning algorithms have to process tremendous amounts of data \citep{lee2001}. The generated models are naturally used for offline anomaly detection, i.e., analyzing consumption data after being loaded into an energy management system (EMS). Effective anomaly detection, however, should happen in real-time in order to minimize the compromises to the use of energy. The efficiency of updating the detection model and the accuracy of the detection are the important consideration for constructing such a real-time anomaly detection system.

\begin{figure}[htp] 
\centering 
\includegraphics[width=0.6\textwidth]{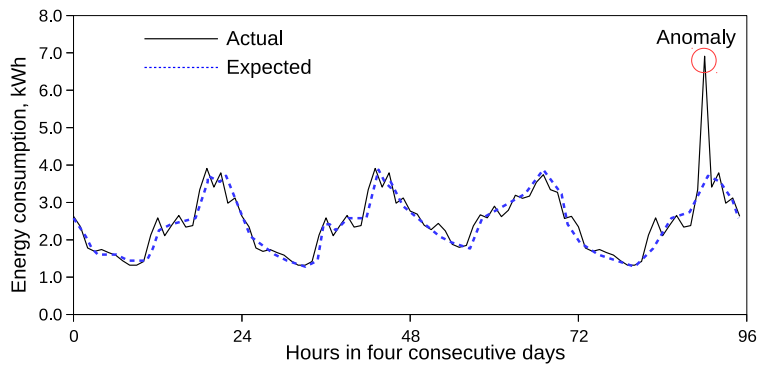} 
\caption{Daily pattern of a typical household and anomaly consumption} 
\label{fig:anomalycase} 
\end{figure}

In this paper, we propose a statistical anomaly detection method based on the consumption patterns of smart grid data. For residential electricity consumption, the daily consumption patterns of a customer usually show quite similar. The proposed algorithm can detect the unusual energy usage from one's history consumption patterns, e.g., the abnormal high usage than the expected (see Fig.~\ref{fig:anomalycase}). 

To detect anomalies in time and obtain a better accuracy, we make use of the so-called {\em Lambda} architecture \citep{Marz2013}, that can detect anomalies in near real-time, and can efficiently update detection models regularly according to a user-specified time interval. A lambda architecture enables real-time updates through a three-layer structure, including speed layer (or real-time layer), batch layer and serving layer. It is a generic system architecture for obtaining near real-time capability, and its three layers use different technologies in processing data. It is well-suited for constructing an anomaly detection system that requires real-time anomaly detection and efficient model refreshment (we will detail it in the next section). To support big data capability, we choose the Spark Streaming as the speed layer technology for detecting anomalies on a large amount of data streams, Hive as the batch layer technology for computing anomaly detection models, and PostgreSQL as the serving layer for saving the models and detected anomalies; and sending feedbacks to customers. The proposed system can be integrated with smart meters for detecting anomalous energy consumption online. To summarize, we make the following contributions: 1) we propose the statistical-based anomaly detection algorithm based on customers' history consumption patterns; 2) we propose making use of the lambda architecture for the efficiency of the model updating and real-time anomaly detection; 3) we implement the system with a lambda architecture using hybrid technologies; 4) we evaluate our system in a cluster environment using realistic data sets, and show the efficiency and effectiveness of using the lambda architecture in a real-time anomaly detection system.

The rest of this paper is organized as follows. Section 2 discusses the anomaly detection algorithm used in the paper. Section 3 describes the implementation of the lambda detection system. Section 4 evaluates the system. Section 5 surveys the related works. Section 6 concludes the paper and provides the direction for the future works.   
\section{Preminaries}

\subsection{Anomaly Detection Model}
The used anomaly detection model is a combination of a short-term energy consumption prediction algorithm, called {\em periodic auto-regression with eXogenous variables (PARX)} \citep{omid}, and {\em Gaussian statistical distribution}. We now first describe the PARX algorithm, which will be used for the prediction based on history consumption patterns. Generally speaking, residential electricity consumption is highly correlated to the weather temperature. In winter, electricity consumption increases since the temperature decrease because of the heating needs. Similarly, in summer, electricity consumption increases when the temperature is higher because of cooling loads. A similar daily consumption pattern may appear repeatedly for a customer, e.g., due to the living habit of the customer. For example, if a customer usually gets up at 7 o'clock, then the consumption pattern will have the morning peak between 7 and 8 o'clock; In the evening, if the customer gets home at 5 o'clock after work, the consumption pattern typically will have the evening peak between 17 and 20 o'clock, due to cooking and washing.

The PARX model, thus, uses a daily period, taking 24 hours of the day as the seasons, \ie, $t=0...23$,  and uses the previous $p$ days' consumptions at the time $t$ for auto-regression. The PARX model at the $s$-th season and at the $n$-th period is formulated as    

\vspace{-10pt}
\begin{equation}
	Y_{s,n} = \sum_{i=1}^{p} \alpha_{s,i} Y_{s,n-i}  + \beta_{s,1} XT1 + \beta_{s,2} XT2 + \beta_{s,3} XT3 	+ \epsilon_s, \enspace s \in t
\label{eq:parx}
\end{equation}
where $Y$ is the data point in the consumption time-series; $p$ is the number of order in the auto-regression; $XT1, XT2$ and $XT3$ are the exogenous variables accounting for the weather temperature, defined in the equations of $(2) - (4)$; $\alpha$ and $\beta$ are the coefficients; and $\epsilon$ is the value of the white noise.   
{\small
\begin{equation}
XT1 = \begin{cases}
T-20 & \text{if } T>20 \\ 
0 & \text{otherwise}  
\end{cases}
\qquad
	XT2 = \begin{cases}
16-T & \text{if } T<16 \\ 
0 & \text{otherwise}  
\end{cases} 
\qquad
	XT3 = \begin{cases}
5-T & \text{if } T<5 \\ 
0 & \text{otherwise}  
\end{cases}
\end{equation}
}

The variables represent the cooling (temperature above 20 degrees), heating 
(temperature below 16 degrees), and overheating (temperature below 5 degrees), respectively. 
The anomaly detection algorithm uses unique variate Gaussian distribution described in the following. 
Given the training data set, $X=\{x_1, x_2, ..., x_n\}$ whose data points obey the normal distribution with the mean $\mu$ and the variance $\delta^2$, the detection function is defined as 
\begin{equation} 
p(x; \mu, \delta)=\frac{1}{\delta \sqrt{2\pi}}e^{-\frac{(x-\mu)^2}{2\delta^2}} 
\end{equation} 
where $\mu=\frac{1}{n} \sum^n_{i=1} x_i$ and $\delta^2=\frac{1}{n}\sum_{i=1}^{n}(x_i-\mu)^2$. For a new data point, $x$, this function computes its probability density. If the probability is less than a user-defined threshold, \ie, $p(x)<\epsilon$, it is classified as an anomaly, otherwise, it is a normal data point. In our model training process, we compute the L1 distance between the actual and predicted consumptions, \ie, $||Y_t - \hat{Y}_t||$, where $Y_t$ is the actual hourly consumption at the time $t$, and $\hat{Y}_t$ is the predicted hourly consumption at the time $t$. The predicted hourly consumption, $\hat{Y}_t$, is computed using the PARX model in Equation~\ref{eq:parx}. We find that the L1 distances obey to a log-normal distribution (see Section~\ref{sec:accuracy}). Therefore, the $x$ in the normal distribution will be the log value of the distance, \ie, $ln||Y_t - \hat{Y}_t||$.

\subsection{Lambda Architecture}
\begin{figure*}[htp] 
\vspace{-10pt}
\centering 
\includegraphics[width=0.9\textwidth]{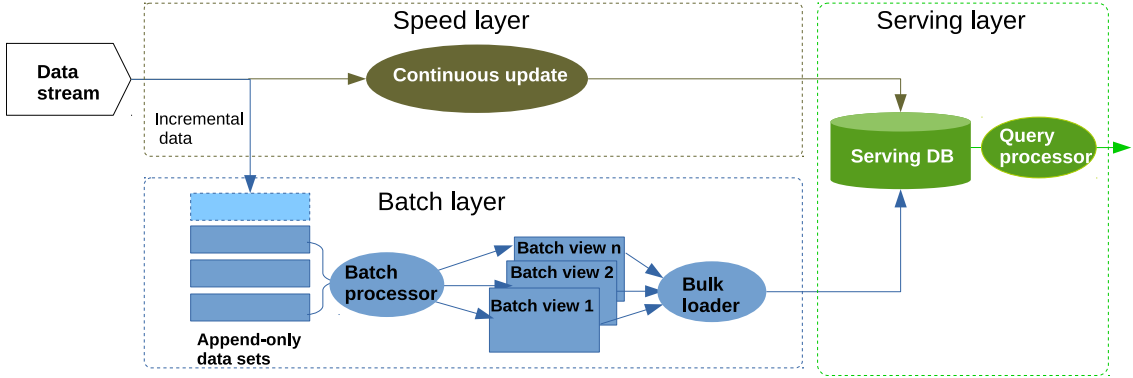} 
\caption{Lambda architecture} 
\label{fig:lambdaarch} 
\vspace{-15pt}
\end{figure*}

We now introduce the Lambda architecture that will be used in our anomaly detection system. As mentioned in Section~\ref{sec:intr}, the lambda architecture consists of three layers, including speed layer, batch layer and serving layer, illustrated in Fig.~\ref{fig:lambdaarch}. The speed layer directly ingests data streams from data sources, processes them, and continuously updates the results into the real-time views in the database in the serving layer. The speed layer does not keep any history records, and typically uses main memory based technologies to analyze the incoming data. In contrast, the batch layer runs iteratively and starts from the beginning of the data set once a batch job has finished. When a batch job starts, all the available data in the batch layer storage will be processed. Therefore, the data arriving after the job starts will not be processed until the next job. Since all the data are analyzed in each iteration, each of the new result views will replace its predecessor. As the batch layer does not rely on incremental processing, it is robust to any system failures, which the batch job simply processes all the available data sets in each iteration. The speed and batch usually use different technologies because of their distinct requirements regarding read and write operations. Any query against the data is answered through the serving layer, \ie, the query processor queries both the views from the speed and the batch layers, and merges them. 

The lambda architecture itself is only a paradigm. The technologies with which the different layers are implemented are independent of the general idea. The speed layer only deals with new data and compensates for the high latency updates of the batch layer. It can typically leverage stream processing systems, such as Storm, S4, and Spark Streaming, etc. The batch layer needs to be horizontally scalable and supporting random reads, where the technologies like Hadoop with Cascading, Scalding, Pig, and Hive, are suitable. The serving layer requires a system with the ability to perform fast random reads and writes. The system can be a high-performance RDBMS (e.g., PostgreSQL), an in-memory data store (\eg, Redis, or Memcache), or a high scalable NoSQL system (\eg, HBase, Cassandra, ElephantDB, MongoDB, or DynamoDB).

\section{\uppercase{Implementation}}

\subsection{System Overview}
\begin{figure}[htp]
\vspace{-15pt}
\centering
\includegraphics[width=0.8\textwidth]{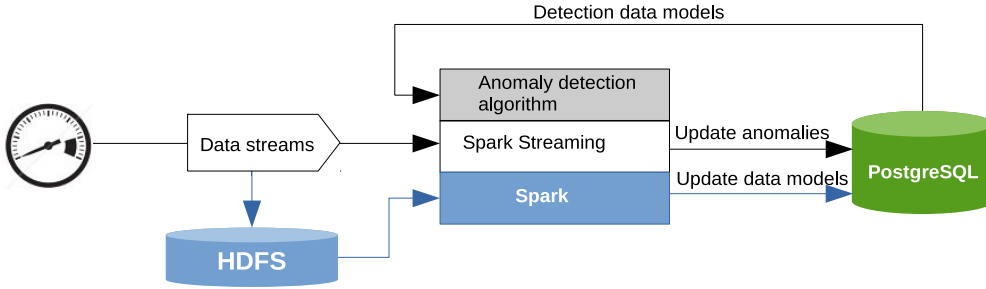}
\vspace{-5pt}
\caption{The anomaly detection system}
\label{fig:systemoverview}
\vspace{-20pt}
\end{figure}

We now describe the implementation of the anomaly detection system. We choose Spark Streaming, Spark, and PostgreSQL as the speed layer, batch layer and serving layer technology, respectively (see Fig.~\ref{fig:systemoverview}). The system employs Spark to compute the models for anomaly detection, which reads the data from the Hadoop distributed file system (HDFS) in the batch layer. The batch job runs at a regular time interval, computes and updates the detection models to the table in PostgreSQL database. Spark Streaming is used for processing real-time data streams, e.g., it directly gets data from smart meters, and detects abnormal consumption by the detection algorithm. The detection algorithm always uses the latest models getting from the PostgreSQL database. Spark Stream writes the detected anomalies back to the PostgreSQL database, which will be used for the notification of customers.   


\subsection{Training Anomaly Detection Models}

\begin{figure}[htp]
\centering
\includegraphics[width=0.7\textwidth]{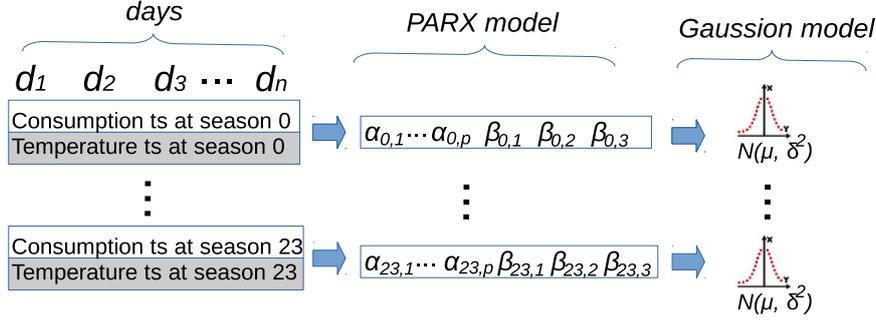}
\caption{Process of training detection models}
\label{fig:training}
\end{figure}

We employ Spark to train the detection model by running regular batch jobs. All the consumption data from smart meters are written to the append-only HDFS. In each iteration of the batch jobs, Spark uses all the available data in HDFS to compute the detection model. The use of Spark and HDFS supports the computation of the model based on scalable data sets, and since they both are the distributed computing technology, the computation can be finished within a certain time limit, which means that the detection algorithm can use the latest data model for the anomaly detection. Fig.~\ref{fig:training} illustrates the training process of generating PARX model and Gaussian model using energy consumption and weather temperature time series at the season from 0 to 23. That is, for each season, e.g., $s=0$, we create a new time series with the hourly reading at 0 o'clock of all days, then use the Equation~\ref{eq:parx} to compute the PARX model (or parameters), and compute the Gaussian model, \ie, $N(\mu, \delta^2)$. Therefore, there are 24 PARX and 24 Gaussian models in total for the hours of the day.   

Algorithm~\ref{alg:training} gives more details about the implementation. This algorithm computes the anomaly detection models with the given training time series collection $\mathcal{TS}$, weather temperature time series $ts'$, and auto-regression order $p$. Each time series in $\mathcal{TS}$ represents the hourly energy consumption of a customer. To compute the detection models for each season $s$, we first need to create a new consumption time series and  a new temperature time series (see line 7), then use the two new time series to compute the PARX model (see line 8). According to our analysis in Section~\ref{sec:accuracy}, the $L1$ distances between predict consumption and actual consumption at season $s$ for all days observes to a log-normal distribution. Therefore, we compute Gaussian statistical model based on the $L1$ distance log values (see line 12-18). The total number of PARX models for all the time series is $||\mathcal{TS}|| \times 24$, which is same as the number of the Gaussian models. In the end, all the models are updated to the PostgreSQL database  that will be used for the online anomaly detection in the speed layer. 
\begin{algorithm}
\caption{Training of anomaly detection models}
{\scriptsize
\begin{algorithmic}[1]
\Function {Train}{TimeSeriesCollection $\mathcal{TS}$, TemperatureTimeSeries $ts'$ Order $p$}
    \State $\mathcal{M} \gets \{ \}$ \Comment{Initialize the collection of PARX parameters}
    \State $\mathcal{N} \gets \{ \}$ \Comment{Initialize the collection of the statical model parameters}
    \ForAll{$ ts \in \mathcal{TS}$ }
        \State $id \gets$ Get the unique identity of $ts$
        \ForAll{$ s \in 0...23$ }
        	\State $ts^c, ts^t \gets$ Construct a new consumption time series using $ts$, and a new temperature time $ts^t$ using $ts'$ at season $s$
		    \State $\alpha_{1},...,\alpha_{p}, \beta_{1},\beta_{2},\beta_{3} \gets$ Compute PARX model using $ts^c$ and $ts^t$
		    \State Insert $(id, s, \alpha_{1},...,\alpha_{p}, \beta_{1},\beta_{2},\beta_{3})$ into $\mathcal{M}$
		     \State $\mathcal{L} \gets \{ \}$
		     \State $\mathcal{D} \gets$ Get the days of $ts$
		    \ForAll{$ d \in \mathcal{D}$ }
		        \State $\hat{v} \gets$ Compute the predict reading of the season $s$ using PARX
		        \State $v \gets$ Get the actual hourly reading from $ts$ of the day $d$
		        \State $l \gets$ Compute the ln value of $L_1$ distance of the day $d$, $ln(||\hat{v} - v||)$
		        \State Add $l$ into  $\mathcal{L}$
		    \EndFor
		    \State $\mu, \beta \gets$ Compute the mean and standard deviation using the normal distribution statistical model on  $\mathcal{L}$
		    \State Insert $(id, s, \mu, \delta )$ into $N$
		    \EndFor
     \EndFor	
    \Return $\mathcal{M}, \mathcal{N}$
 \EndFunction
\end{algorithmic}
}
\label{alg:training}
\end{algorithm}

The implementation is a Spark program. The consumption time series, as well as temperature time series, are read into the distributed main memory as {\em resilient distributed datasets (RDDs)}, which are fault-tolerant, immutable and partitioned parallel data structures that can be operated in parallel, e.g., by using the operators, including map, reduce, groupByKey, filter, collect, etc \citep{Zaharia}. To generate the new time series, we use the {\em groupByKey} operator to aggregate the consumption series by the composite key of meter ID and season (or hours); while use only the season as the key to the temperature time series. Then, we merge the generated time series by the \texttt{join} operator on the key of the season. The PARX, in fact, can be regarded as a multi-linear regression model, which simply takes the auto-regressors and the exogenous variables as the independent variables. We, then, apply the multiple linear regression function from the Spark machine learning the library, MLib \citep{Meng}, to compute the coefficients. For all of these operations, the transformation functions are directly applied on RDDs for doing the data processing.    

\subsection{Real-time Anomaly Detection}
The real-time anomaly detection is carried out in the speed layer. Algorithm~\ref{alg:realtimedetection} describes the anomaly detection process, which is self-explanatory. First, the detection algorithm reads meter readings from all the incoming data streams, and reads the weather temperature and detection models from the PostgreSQL database each hour. For each data stream, the algorithm predicts the reading using the PARX algorithm, with the pre-computed parameters, the previous $p$ day's readings at the current hour, \ie, the season $s$, and weather temperature (see line 4--7). Then, the algorithm calculates the log value of the $L1$ distance between the predict and the actual readings, then uses it compute the probability using the Gaussian model (see line 8-10). In the end, the algorithm decides whether the current reading is an anomaly or not based on the computed probability value, \ie, if its value is below the user-defined threshold, $\epsilon$. If the current reading is classified as an anomaly, it will be written into the database for the customer notification (see line 11--12).
\begin{algorithm}[htp]
\caption{Real-time anomaly detection}
{\scriptsize
\begin{algorithmic}[1]
\Function {Detect}{CurrentReadingCollection $\mathcal{V}$, Temperature $t$, PredictModel $\mathcal{M}$, StaticalModel $\mathcal{N}$, Threshold $\epsilon$ }
    \State $\mathcal{R} \gets \{ \}$ \Comment{Initialize the detection results}
    \ForAll{$ v \in \mathcal{V}$ }
        \State $id \gets$ Get the unique identity of $v$
        \State $s \gets$ Get the season of $v$
        \State $\alpha_{1},...,\alpha_{p}, \beta_{1},\beta_{2},\beta_{3} \gets$ Get the parameters from  $\mathcal{M}$ by $id$
        \State $\hat{v} \gets$ Compute the predict reading at $s$ using PARX with the parameters, the $p$ days' readings at $s$, and temperature $t$
        \State $x \gets ln||\hat{v} - v||$ \Comment{Compute the ln value of $L_1$ distance at the season $s$}
        \State $\mu, \delta \gets$ Get the statical model parameters from $\mathcal{N}$ by $id$ and $s$
        \State $p \gets$ Compute the probability using the normal distribution function, $\frac{1}{\delta \sqrt{2\pi}}e^{-\frac{(x-\mu)^2}{2\delta^2}}$
        \If {$p<\epsilon$}
         \State  Add $(id, s, p, v, \hat{v})$ into $\mathcal{R}$
        \EndIf
     \EndFor	
    \Return $\mathcal{R}$
 \EndFunction
\end{algorithmic}
}
\label{alg:realtimedetection}
\end{algorithm}

\begin{figure}
\centering
\includegraphics[width=0.85\textwidth]{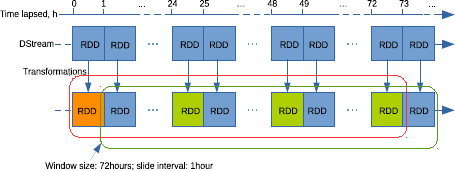}
\vspace{-5pt}
\caption{Slide windows in the real-time anomaly detection}
\label{fig:rddslidewin}
\end{figure}

We implement the algorithm to process the real-time data on Spark Streaming. Spark Streaming allows for continuous processing via short interval batches, and its basic data abstraction is called {\em discretized streams (D-Streams)}, a continuous stream of data \citep{sparkstreaming}. The data are received in each interval batch, {\em hourly} in our case, and operations will run upon the data for doing transformations, such as filter unnecessary attribute values, extracting the hour from the timestamp, etc. (see Fig.~\ref{fig:rddslidewin}). When using PARX for the prediction, we fetch the previous $p$ days' readings of the current hour for auto-regression. For example, in Fig.~\ref{fig:rddslidewin} we set the order, $p=3$, therefore, the window size is set to 72 hours (\ie, 3 days)  to keep the past three days' readings at the particular hour within the same window (e.g., the RDDs colored by green). This is done by using the window function, \texttt{reduceByKeyAndWindow(func, windowLength, slideInterval)}, to aggregate the data with specified key, window length and slide interval (\eg, meter ID and season as the composite key in this case,  $windowLength=72$ hours and $slideInterval=1$ hour). In the underlying, Spark uses  the data {\em checkpointing} mechanism to keep the past RDDs in HDFS. At the beginning of each interval batch, the data models are read from the PostgreSQL database in the serving layer, and broadcast to all DStreams. Therefore, the detection program can always use the latest detection models to detect anomalies.   

\section{\uppercase{Evaluation}}
\subsection{Experimental settings}
In this section, we will evaluate the effectiveness and the scalability of our anomaly detection system. We conduct the experiments in a cluster with 17 servers.  Five servers
are used for running the speed layer, while twelve servers are used for the batch layer.
We also exploit one of the servers in the speed layer as the serving layer for managing
the detection models and sending anomaly detection messages. All the servers have the identical settings, configured with an Intel(R) Core(TM) i7-4770 processor (3.40GHz, 4 Cores, hyper-threading is enabled, two hyper-threads per core), 16GB RAM, and a Seagate Hard driver (1TB, 6 GB/s, 32 MB Cache and 7200 RPM), running Ubuntu 12.04 LTS with 64bit Linux 3.11.0 kernel. The serving layer uses PostgreSQL 9.4 database with the settings ``shared buffers=4096MB, temp buffers=512MB, work mem=1024MB, checkpoint segments=64'' and default values for the other configuration parameters.

We have a real-world residential electricity consumption data set (27,300 time series), which will be used for evaluating the accuracy of the anomaly detection. The data are the 2-year time series of hourly resolution. To evaluate the scalability, we use the synthetic data set generated by our data generator with the real-world data as the seed. The size of data tested in the cluster environment is scaled up to one terabyte, corresponding to over twenty million time series. 

\subsection{Anomaly Detection Accuracy}
\label{sec:accuracy}
We start by evaluating the accuracy of our anomaly detection system using a randomly-selected time series from the real-world data set. 

\begin{figure}[htp]
	\vspace{-10pt}
\begin{minipage}[b]{0.4\textwidth}
	\centering
	\includegraphics[width=1.35\textwidth]{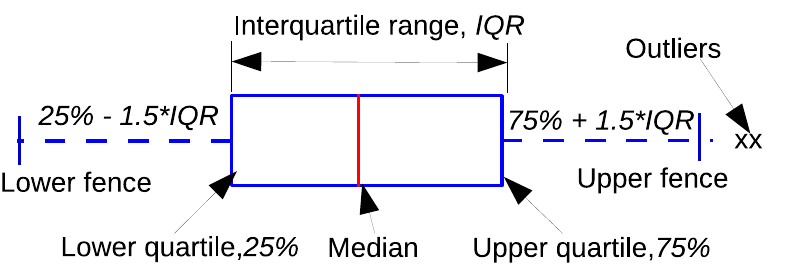}
	\vspace{-18pt}
	\caption{Box plot}
	\label{fig:boxplot}
\end{minipage}
\begin{minipage}[b]{0.75\textwidth}
		\centering
		\includegraphics[width=0.56\textwidth]{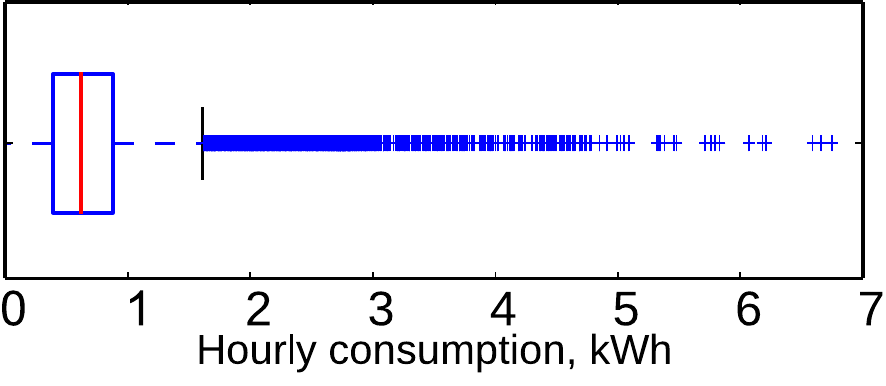}
		\vspace{-10pt}
		\caption{Anomaly detection using box plot}
		\label{fig:anomalyboxplot}
	   \end{minipage}
\vspace{-20pt}
\end{figure}

To provide a basis for comparison, we perform the anomaly detection using a standard boxplot analysis as well. Boxplot is a quick graphic approach for examining data sets, and has been used for decades. A boxplot uses five parameters to describe a numeric data set, including lower fence, lower quartile, media, upper quartile and upper fence (see Fig.~\ref{fig:boxplot}). According to Fig.~\ref{fig:boxplot}, a boxplot is constructed by drawing a rectangle between the upper and lower quartiles with a solid line indicating the median. The length of the box is called interquartile range, $IQR$. The sample data points lying outside the fences, $1.5*IQR$, are classified as the outliers, which has been indicated to be acceptable for most situations \citep{frigge1989}.

To align boxplot with our detection method, we test the anomalies based on the 24 seasons. There are 17,520 data points in total in the selected time series. Fig.~\ref{fig:anomalyboxplot} shows the boxplot result where the blue points located on the top of the upper fence represent the anomalies, a total of 1,260 data points. Since the boxplot approach is merely able to detect energy consumption lying unusually far from the main body of the data,  it is difficult to determine which ones are the true anomalies, and to identify the potential reasons for these anomalies because there are too many false positives.

\begin{figure}[t]
\vspace{-5pt}
  \begin{minipage}[b]{0.33\textwidth}
    \centering
	\includegraphics[width=1.02\textwidth]{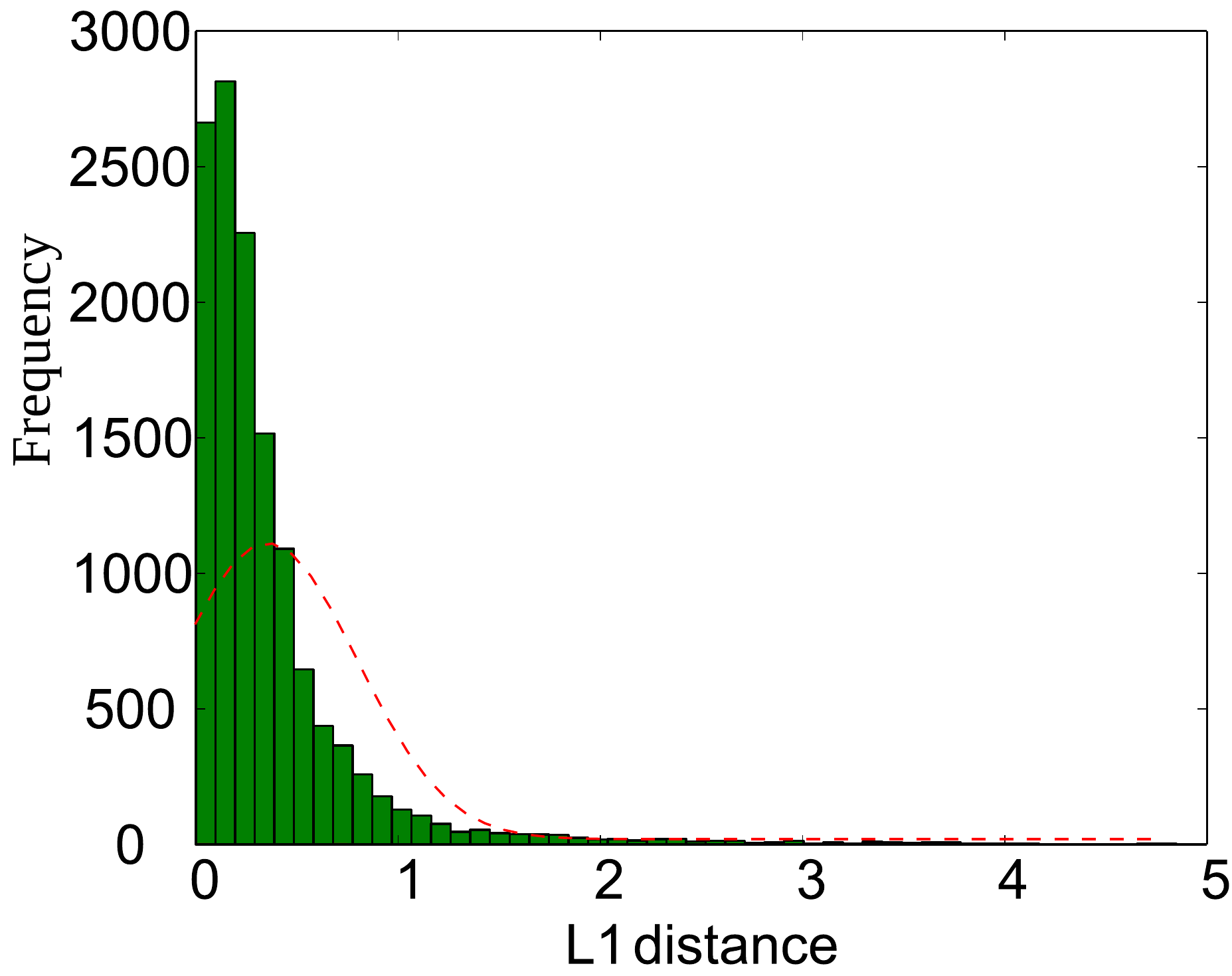}
	\vspace{-20pt}
	\caption{Log-normal distributions across the L1 distances }
	\label{fig:l1distancedist}
  \end{minipage}
  \begin{minipage}[b]{0.33\textwidth}
    \centering
	\includegraphics[width=0.9\textwidth]{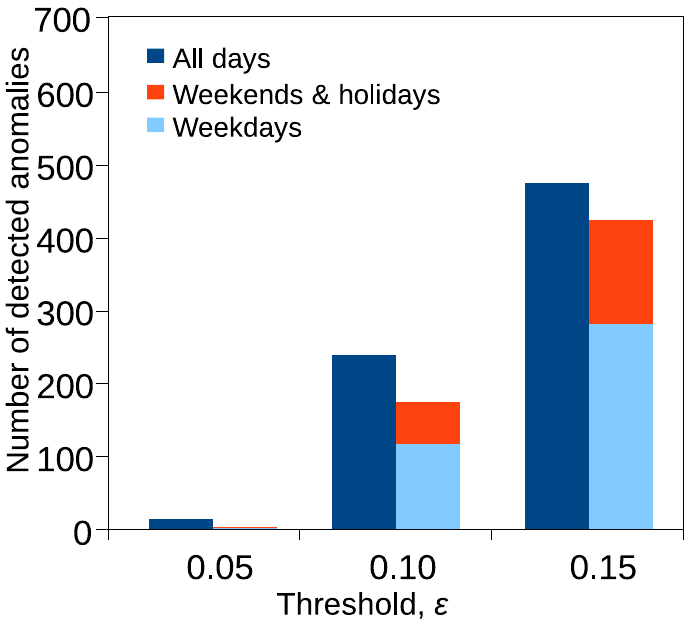}
	\vspace{-10pt}
	\caption{Anomaly detection using PARX and statistical method}
	\label{fig:detectedanomalies}
  \end{minipage}
   \begin{minipage}[b]{0.33\textwidth}
     \centering
	\includegraphics[width=0.9\textwidth]{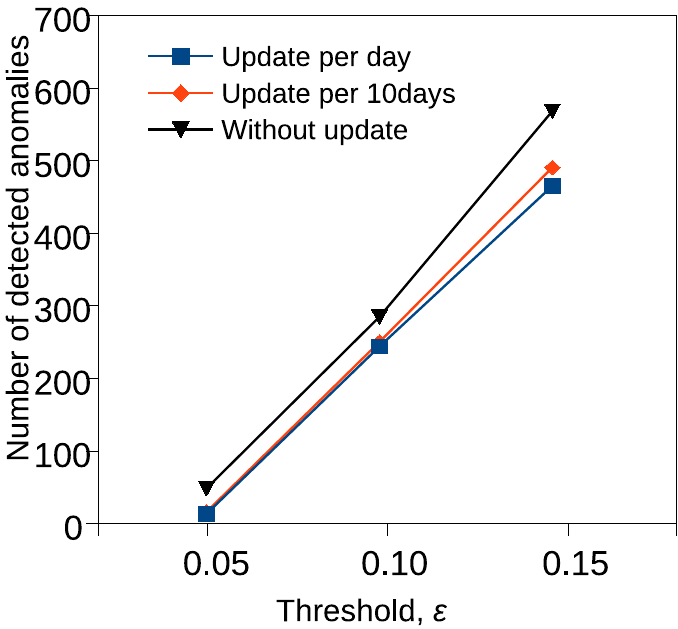}
	\vspace{-10pt}
	\caption{Impact of detection model update frequency}
	\label{fig:detectmodelupdate}
  \end{minipage}
  \vspace{-10pt}
\end{figure}

We now use the proposed detection algorithm to analyze the same time series. Fig.~\ref{fig:l1distancedist} depicts the distribution of the $L1$ distances of a season by the histogram. As shown, the distribution has the shape of a log-normal distribution. We have checked the $L1$ distance distributions for all the 24 seasons, and found that they all have a  similar shape. This is the reason that we choose log-normal distribution in our statistical-based anomaly detection. Besides, we test the anomalies by treating all the days the same, and different, \ie, discriminating the days into workdays, weekend \& holidays.  Moreover, we increase the threshold value, $\epsilon$, from 0.05 to 0.15, and do the test. The results in Fig.~\ref{fig:detectedanomalies} demonstrate that the detection identifies more anomalies for treating all the days the same than different. The reason is that during weekends and holidays, people tend to stay at home more time, thus use more energy. The consumptions are more likely higher than in the weekdays. For the threshold parameter, its value is for classifying a usual or unusual reading. According to the results, if the value increases, the number of detected anomalies changes significantly. For the real-world deployment of this system, the threshold value can be set by the residents to decide when to receive anomaly alerting messages.


We now evaluate the impact of the model update frequency on the detection accuracy. We use half-year's time series as the initial data set to train the detection models. We design the following three scenarios for model updating: 1) update per day; 2) update per 10 days behind the detection; and 3) without an update. We measure the detected anomalies for the three scenarios by treating all days the same. According to the results shown in Fig.~\ref{fig:detectmodelupdate}, the frequent updates of the models help to decrease the detected anomalies. It is due to the improvement of the prediction accuracy of the PARX model. Thus, less large $L1$ distances are identified as the anomalies. However, although the update frequency does help to determine the real anomalies, the results do not show a big difference if the models are updated within a certain short-time interval, \eg, the results of the scenario 1) and 2).

In the end, we compare our approach with the boxplot, and the result shows that the number of the anomalies reported by the PARX prediction and statistical method can be decreased notably. This increases the chance to determine accurately real anomalies for an energy consumption time series.

\subsection{System Scalability}
We now evaluate the scalability of our anomaly detection system. As our system can scale-out and to efficiently cope with large amounts of data, we vary both the number of executors and the volume of the input data in the following.

\begin{figure*}[t]
  \begin{minipage}[b]{0.33\textwidth}
  \centering
	\includegraphics[width=0.95\textwidth]{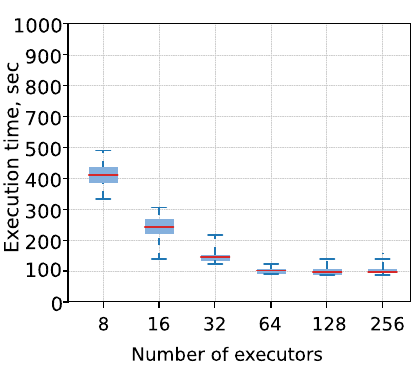}
	\caption{Batch model training}
	\label{fig:scaleoutbatch}
  \end{minipage}
  \begin{minipage}[b]{0.33\textwidth}
    \centering
	\includegraphics[width=0.95\textwidth]{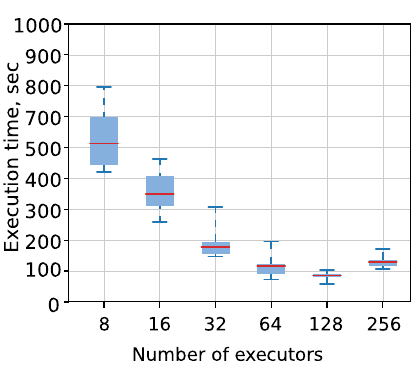}
	\caption{Real-time detection}
	\label{fig:scaleoutstream}
	\end{minipage}
  \begin{minipage}[b]{0.33\textwidth}
    \centering
	\includegraphics[width=1\textwidth]{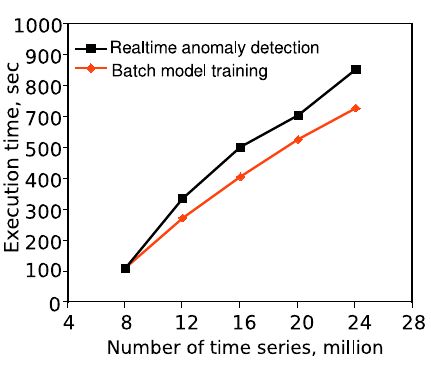}
	\caption{Size-up experiment}
	\label{fig:sizeup}
	\end{minipage}
\end{figure*}

{\bf Scale-out experiment.} Parallel processing is a key feature of the proposed system. To evaluate the scalability of our implementation, we conduct an experiment with a varying number of the executors in Spark. In this experiment, we use a fix-sized synthetic data set with eight million time series of a one-year length (275GB), which were generated by our data generator seeded with the real-world data set. Since we are interested in the real-time and batch capability of our system, we use the full sets of the nodes to test  batch model training and real-time anomaly detection separately. We first test the batch capability by running the training program with the number of executors increased from 8 to 256. We repeat each test for ten times, and record their execution times. The results are depicted by the boxplot shown in Fig.~\ref{fig:scaleoutbatch}. According to the results, the execution time and the time variance decrease when more executors are added. But, when the number reaches 64, the increasing parallelism does not speed up the batch processing further, which is due to the overhead of the Spark master when managing a large number of executors. We conduct the real-time anomaly detection on Spark Streaming, and likewise, we scale the number of executors from 8 to 256. Since we only study the real-time detection scalability, we use the detection models without updating. Fig.~\ref{fig:scaleoutstream} shows the results, which indicate that the variance of execution time is larger than the batch model training. It might be due to the variability of real-time batch executions on Spark Streaming when doing the anomaly detection for each hour.

{\bf Increased data load experiment.}  To evaluate the scalability of the algorithms over large volumes of data, we compare different workloads. According to the above experiments, the optimal number of parallel executors for model training and anomaly detection are 64 and 128, respectively. We use the optimal executor number (the memory of executor is configured to 4GB) in our experiments, but vary the number of time series from 8 to 24 million (the size is from 275GB to 825GB). The processing times of varying data workloads are displayed in Fig.~\ref{fig:sizeup}. We observe that both of the training and detection processes can scale near linearly with the quantity of the time series. The time of detection, in this case, is the total execution time of handling all the time series of a one-year length, \eg, it takes less than two minutes to finish eight million time series with the optimized settings. The average time of each real-time batch only takes a few seconds (recall that a batch in Spark Streaming processes the data of each hour). In the real-world deployment, the detection program can be set to run every hour to inspect hourly smart meter readings. According to the results, the anomaly detection system has a very good scalability which can meet the fast-growth of smart meter data.

Unlike the anomaly detection, the training process uses the full set of the data to generate new models each time.  Anomaly detection, instead, is performed for each hour where Spark Streaming runs periodical batch (or impulse) to process the data, which needs more time in overall. The training and detection programs can be deployed either in different clusters or the same cluster. If deployed in the same cluster, it is necessary to allocate computing resources in a reasonable way. For example, since the batch job takes a much longer time, it can be scheduled to run immediately after the anomaly detection job. A scheduling system is, thus, necessary, and this will be for our future work.

\section{\uppercase{Related Works}}
{\bf Anomaly energy consumption detection.} 
Anomaly detection is an important aspect in energy consumption time series management. Chandola \et present a survey of different anomaly detection techniques in various application domains including energy \citep{Chandola2009}. Statistical and data mining are the commonly used techniques for discovering abnormal consumption behaviors \citep{Janetzko}. Statistical methods are based on modeling data using distributions, and see if the data under test observes to the distributions. Accordingly, the approaches presented in this paper combine PARX and log-normal distribution function to detect anomalies in energy consumption time series. Jakkula and Cook use statistics and clustering to identify outliers in power datasets collected from smart environments \citep{Jakkula2010}, but they have not considered the impact of the exogenous variables, \eg, weather temperature,  on the electricity consumption. Linear regression can extract time series features when the dependent variables are well-defined \citep{Magld2012}. The early experience of identifying outliers in linear regression is through setting a threshold limit, but this yields many false positives for large data sets \citep{lee1997}. Adnan \et combine linear regression with clustering techniques for getting better results \citep{Adnan2003}. Zhang \et \citep{zhang2011} further  use piecewise linear regression to fit the relation between energy consumption and weather temperature. The results obtained are more favorable than entropy and clustering methods. But, their approach does not take the changes of consumption pattern into account. Brown \et use K-nearest neighborhood (KNN) in fast kernel regression to predict electricity consumption \citep{brown2011}, which requires large datasets. The resulting models are static, thus it is not preferable for online anomaly detection and the situation when consumption pattern is changed. Nadai \et combine ARIMA and adaptive artificial neural network (ANN) to detect anomaly consumption \citep{nadai2015} using a relatively small data set that is from a few buildings. In comparison, we propose the prediction and statistical anomaly models and combine with the lambda architecture for supporting regular model refreshment, and real-time anomaly detections. Besides, the proposed approach can handle scalable data sets, and consumption pattern changing owing to using the PARX model.

{\bf Batch and stream processing on big data.} 
Batch and realtime/stream processings have attracted much research effort in recent year, with the popularity of Internet of Things (IoT). Liu \et make a survey of the existing stream processing systems, and discuss the potential technologies used for lambda architecture \citep{Liu2014}. The tools \citep{dawak2011,tldks2011,pvldb2011,ideas2014} Cheng \et propose a smart city data platform that supports both batch and real-time data processing \citep{Cheng}, and they suggest that anomaly detection should be implemented as the chief component of any platform for processing sensor data. Different to proposing the generic lambda architecture  \citep{Marz2013}, Preuveneers \et \citep{Preuveneers} and Gao \citep{Gao} present the big data architectures for processing domain-specific big data, including health care, context-aware user authentication and social media.  Schneider \et study batch data and streaming data anomaly detection, respectively \citep{Schneider2016}. The used detection model, however, is static, and the use case is different to ours which employs the batch job to update the models iteratively while use the real-time job to detect the anomalies online in data streams.  

{\bf The use of lambda architecture.}
Lambda architecture has attracted a growing interest due to its mix capabilities to process both real-time and batch data. Sequeira \et use lambda architecture in an industrial EMS solution with cloud computing capabilities \citep{Sequeira2014}. Kro\ss \et develop on-demand stream processing within the lambda architecture to optimize the usage of computing resources in clusters \citep{Brunnert}. Martnez-Prieto \et adapts the lambda architecture in semantic data processing \citep{Martnez2015}; Liu \et applies it to smart grid complex event processing (CEP) \citep{Liu2015}; Villari \et proposes AllJoyn Lambda, the platform for managing embedded devices of smart homes \citep{Villari}; and Hasani \et use it for real-time big data analytics \citep{Hasani}. Besides, the works \citep{Casado2015,Liu2014} both give an extensive review of the technologies for the lambda architecture. There are various use cases of the lambda architecture. In contrast, we focus on a particular use case, \ie, energy consumption anomaly detection as has been mentioned in future extensions of \citep{Kiran} work. More specifically, we use it in the model update and the real-time anomaly detection using the models, which is significant to the large deployment of smart meters and sensors of IoT today.

\section{\uppercase{Conclusions and Future Work}}
\label{sec:conclusion}
Analyzing and detecting anomalies is an important task for live energy consumption data while the improvement of detection accuracy and scalability is quite challenging. In this paper, we applied the novel lambda architecture technique to an anomaly detection system, which supports iterative batch updates of the detection models, and real-time anomaly detection on scalable data streams. We have proposed the detection algorithm for finding the anomalies based on one's history consumption pattern via the supervised learning and statistical algorithm. Furthermore, the system supports alerting service for customers by setting a personalized threshold value for conspicuous energy consumption. We have evaluated the accuracy of the anomaly detection algorithm using a real-world data set, and the scalability using a large synthetic data set. The results have validated the effectiveness and the efficiency of the proposed system with the lambda architecture.

For the future work, we will implement a scheduling system that can coordinate the running of the batch and real-time jobs within the same cluster. We intend to explore the ways to detect a greater range of anomalies, such as missing values, negative energy consumption and device errors. Besides, we plan to support additional types of data, such as gas, heating and water data, and to implement the corresponding detection algorithm.

\section*{Acknowledgements}
\noindent This work is part of the CITIES (NO. 1035-00027B) research project funded by Innovation Found Denmark.

\end{document}